\documentclass[12pt]{iopart}
\usepackage{graphicx}

\begin{document}

\title{ Liquid polymorphism, density anomaly and H-bond disruption in an associating lattice gases }  

\author{Aline Lopes Balladares}
  
\address{ Instituto de Fis\'ica, Universidade Federal do Rio Grande 
do Sul \\ 
Caixa Postal 15051, 91501-970, Porto Alegre,RS, Brazil }

\author{Vera B. Henriques}
  
\address{Instituto de F\'{\i}sica, Universidade de S\~ao Paulo,
Caixa Postal 66318, 05315970, S\~ao Paulo, SP, Brazil }

\author{Marcia C. Barbosa\footnote[3] {To whom correspondence should be 
addressed (marcia.barbosa@ufrgs.br)}}

\address{Instituto de Fis\'ica, Universidade Federal do Rio Grande do
 Sul \\ Caixa Postal 15051, 91501-970, Porto Alegre,RS, Brazil}

\begin{abstract}

We have investigated the effects of either distorting hydrogen bonds or removing proton degeneracy
on the thermodynamic properties of a minimal model for associating liquids. The presence of
two liquid phases and a density anomaly is unaffected in both cases. Increasing the
degeneracy of bonded structures leads to lower temperature critical points and
a steeper liquid-liquid coexistence line,
implying a low density liquid of larger entropy. 

Analysis of the hydrogen bond net accross the
phase diagram indicates that the density anomaly is accompanied by a steep reduction of hydrogen
bond density, which introduces a restriction on a correlation which has been preconized long ago.
 This feature is present independent of bond distortion or of the presence of proton entropy.

\end{abstract}

\maketitle

\section{Introduction}

Network-forming liquids such as water are ubiquitous \cite{De98} in nature.
They differ from normal liquids by the presence of 
directional intermolecular interactions that result in
the formation of bonds. These directional attractive
forces favor the formation of structured regions that,
due to the orientational constraints on the bonded 
molecules, have lower density than unbonded regions.
As a result, a density anomaly, consisting in the 
expanion under isobaric 
cooling of these systems, appears. This density anomaly
has been related to a phase transition between
a low-density liquid (LDL) and a high-density liquid (HDL).
Experiments and simulations of water predict a HDL-LDL first-order
 phase transition in an experimentally inacessible region of 
the phase diagram \cite{Po92,Ta96,Mi00,Fr03,Sc03}. 
But water is not an isolated case, computer simulations of
realistic models for carbon \cite{Gl99}, phosphorus \cite{Mo03}, $SiO_2$ 
\cite{Sa01}, 
and Si \cite{An96,Sa03} suggest the existence  first-order LDL-HDL phase
transitions in these materials. 

The presence of a number of solid phases in water, as well as of 
solid-solid
first-order phase transitions, have suggested the possibility 
that systems with solid polymorphism exhibit several liquid
phases with local structures similar to the ones present
in the crystal phases. This assumption was confirmed 
for  a number polymorfic liquids such as Se, S, Bi , P, $I_2$, Sn, Sb,
$As_2Se_3$, $As_2S_3$ and $Mg_3Bi_2$ \cite{Br02,Br98}.

In all these cases, a full understanding of the effects of the number,
spatial orientation and strength of the bonds is still missing. 
In order to shed some light onto this problem, recently 
a simple approach  representing  hydrogen bonds through ice
variables\cite{Hu83,At96,Na91,Gu00} has
been proposed. The zero temperature ice model was succesful in giving 
the description of ice \cite{Li67} entropy, for dense systems 
but an
order-disorder transition for finite temperatures is absent. Recently  a description 
based also on ice
variables but which allows for a low density ordered 
structure \cite{He05a,He05b} was proposed. The associating lattice gas 
 model (ALG)\cite{He05a,He05b}
is based on the
competition between the filling up of the lattice and the formation of an 
open four-bonded orientational
structure which is naturally introduced in terms of the ice bonding variables and
no \emph{ad hoc} introduction of density or bond strength variations is
needed. In our previous 
publications, we have shown that 
this model is able to exhibit, for a convenient set of parameters, both 
density anomalies and 
the two liquid phases \cite{He05a}. It was also shown that by 
varying the relative strength
of the orientational interaction, we can go at a fixed temperature 
from two coexisting liquid phases, as
observed in amorphous water, to a smooth transition between two 
amorphous structures, 
as might be the case of silica \cite{He05b}.

In the case of water, the region of the pressure vs. temperature (p-T) phase
diagram where the density anomaly is present, the preence of distorted 
hydrogen bonds 
which favor the presence of intersticial nonbonded water 
molecules has been reported \cite{Ne01}. These 
nonbonded molecules weaken the actual bonded interaction
in their vicinity. 

On the other hand, investigations on orientational models without distinction between 
donor and acceptor arms \cite{Pa99}
pose the question of whether this distinction,
essential in the case of the original ice entropy problem \cite{Li67},
has any effect on some of the important anomalous properties one aims
to represent.

In this study we firstly wish to contribute to the search for a minimum model for anomalous
associating liquids, in establishing which of the microscopic properties are essential
in order to reproduce the macroscopic behaviour expected for water. 
We propose to analyse two features: 
(i) the presence of distorted bonds and (ii) the absence of proton entropy on bonds.

A second purpose of our work is to look for the correlations between
hydrogen bond behaviour and the density anomaly. 
The water density anomaly has for a long time been associated with bond-breaking or bond-distortion 
\cite{eisenberg}. As temperature is increased, bonds break or distort and allow an increased 
number of neighbours per molecule. As temperature increases further, usual translation
 entropy dictates dilution, and the more usual reduction in density prevails. However,
it is well known that the TMD line (line of maximum density) is restricted to some range
 of pressures, in the
vicinity of the liquid-liquid line. This means that pressure is required to play a role
in inducing an increase in density as temperature rises. In this study we also aim at
analysing the relation between bond-breaking and rising density as functions of temperature.

The remainder of the paper goes as follows. In sec. II, the model for 
distorted hydrogen
bonds is introduced and the corresponding phase diagrams are obtained. In sec. III, a 
simplified version of
the original associating lattice gas model in which the disctinction between donors and receptors 
is removed is presented and its properties 
compared with those of the original assymmetric model. Results for the hydrogen-bond densities
 are presented in
section IV. Finally, conclusions are presented in sec. V.

\section{The distorted-bond associating lattice gas model}

Consider a two-dimensional
triangular lattice where each site may be empty or full. Associate to 
each site two 
kinds of variables: occupational variables, $\sigma_{i}$, and
orientational variables, $\tau_{i}^{ij}$. The orientational state of 
particle $i$ is defined by
the configuration of its bonding and non-bonding arms, as illustrated 
in Fig 1. We consider
three possible values
for the $ij$ arm variables. Four 
are the usual ice bonding arms, two donor ($\tau_{i}^{ij}$=1) and two 
acceptor ($\tau_{i}^{ij}$=-1), and two additional arms are taken as 
inert or non-bonding ($\tau_{i}^{ij}$=0). A bond is formed if a donor 
arm points to a
nearest neighbor acceptor arm. Bonding arms can be 'proper' or 
'distorted'. Proper bonding is considered for the cases in which
the non-bonding arms are opposite, as in Fig. 1a. Bonds may be 
distorted if the non-bonding
arms make an angle of 120 degrees, as shown in Fig. 1a. Thus there 
are three proper bonding
states and six distorted bonding states per particle, making up 
fifty four possible states
 per occupied site.
\begin{figure}
\begin{center}
\includegraphics[width=4cm,height=4cm]{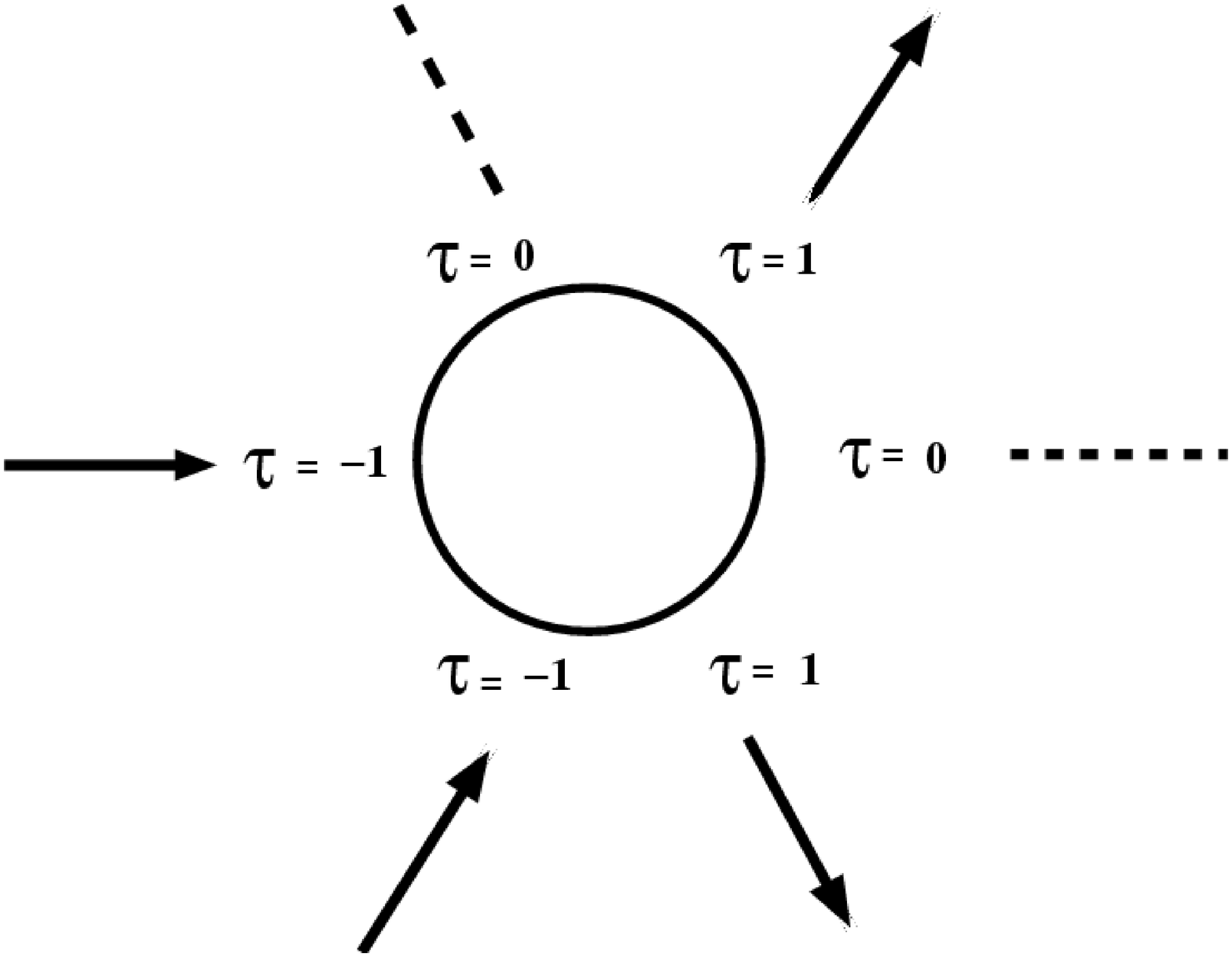}
\hspace{0.3cm}
\includegraphics[width=4cm,height=4cm]{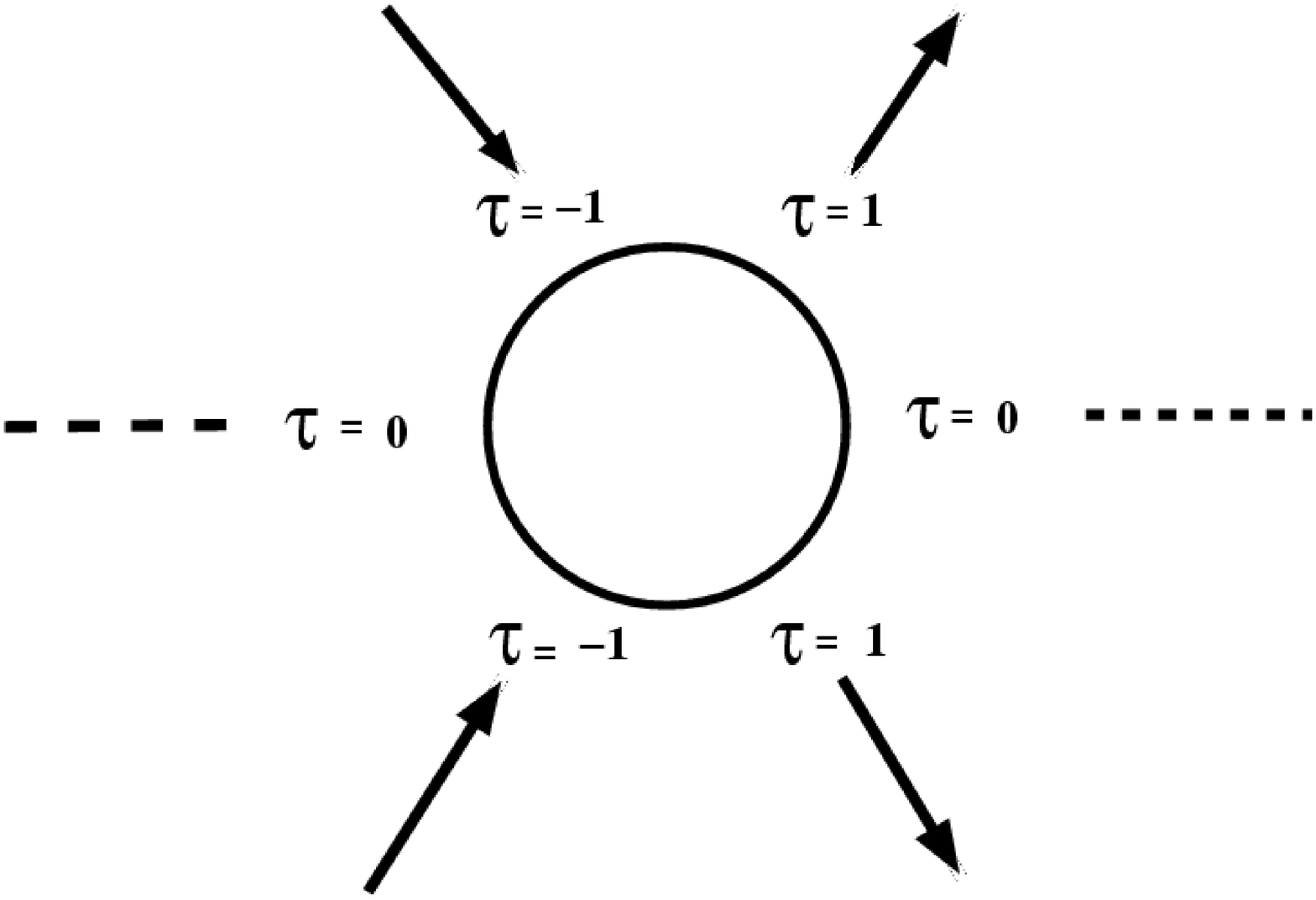}

($a$) \hspace{4cm} ($b$)

\caption{The model orientational state: four bonding 
(donor and receptor) and $(a)$ two distorted non-bonding 
arms and $(b)$ two opposite arms.}
\end{center}
\label{fig1}
\end{figure}

An energy $-v$ is attributed to each pair of occupied neighboring 
sites that form a hydrogen bond in case neither molecule has distorted arms,
while non-bonding pairs are attributed an 
energy of $-v+2u_{1}$. If at least one of the two molecules
has a distorted bonding arm, the energy of the pair is given by $-v+2u_1-2u_2$.
 This makes  $-2u_1$ 
the energy of proper hydrogen bonds whereas $-2u_2$ is the energy per 
bond of molecules
 in the distorted local net. The penalty for distortion is thus $2(u_1-u_2)$
 
 The 
overall model energy is given by
\begin{equation}
\label{E1}
E=(-v+2u_{1})\sum _{(i,j)}\sigma _{i}\sigma _{j} + \sum _{(i,j)} u_{i,j}
 \sigma _{i}\sigma _{j}\tau ^{ij}_{i}\tau ^{ji}_{j}(1-\tau ^{ij}_{i}
\tau ^{ji}_{j})
\end{equation}
where, $\sigma_{i}=0,1$ are the occupational variables, 
$\tau_{i}^{i,j}=0,\pm 1$ represents
the arm states described above. As for H-bonds, described by the 
parameters $u_{i,j}$, 
we have $u_{i,j}=u_1$ in case
both molecules at sites $i$ and $j$ have opposite non-bonding arms, whereas 
$u_{i,j}=u_2$, if at least one of the two molecules $i$  and
$j$ are distorted. Note that each particle 
may have six neighbors, but the number of bonds per molecule 
is limited to four.

This system can exhibit a number of ordered states. Two of 
them, without distorted bonds, are illustrated in Fig. 2. In Figure 2a
 a system fully occupied with each molecule making four hydrogen
bonds is shown. This is the high density liquid phase. 
Figure 2b illustrates the configuration in which
the system has $3/4$ of its sites occupied and each
site has four hydrogen bonds. This is the 
low density liquid phase.  In both cases, none of
the molecules is distorted and the energies per site are
given by $e=-3v+2u_1 $ and $e=-3v/2$ respectively.  Another low 
energy configuration, with distortions, is illustrated in Fig. 3. In 
this case, the system is fully occupied with
all molecules making four hydrogen bonds, however, differently
from the configuration in figure 2a, all molecules
are distorted and the energy per site is  $e=-3v+6u_{1}-4u_{2}$.

At zero temperature, the phase diagram is obtained simply
by comparing the grand potential  per site of the different
configurations.  Here we will assume that $u_1>u_2$ so that
the distortion is punished by having a 
higher energy. Thus, at high chemical potential, the lowest 
grand potential per site is the one of the high density liquid 
phase without distortions ($\rho=1$), 
illustrated in Figure 2a, $\phi_{hdl}=-3v+2u_1-\mu$. As the 
chemical potential
is decreased, the low density liquid ($\rho=0.75$) without distortions with the grand potential per site given by 
$\phi_{ldl}=-3v/2-3\mu/4$ becomes
energetically more favorable and,  at $\mu=-6v+8u_1$, there is 
a transition between 
a high density liquid and a low density liquid. Similarly the 
pressure of coexistence between 
the two liquid phases at zero temperature is given by $p=-3v+6u_1$.

If the chemical potential decreases even further the gas phase 
with $\phi_{gas}=o$ becomes energetically more favorable and 
at $\mu=-2v$ and $p=0$ there is a phase transition between a low 
density liquid
phase and a gas phase. The condition for the presence of the two 
liquid phases is therefore 
$u_{1}/v>0.5$. For lower values of $u_{1}/v>0.5$, the LDL disappears. 

\begin{figure}
\begin{center}
\includegraphics[height=7cm,width=7cm]{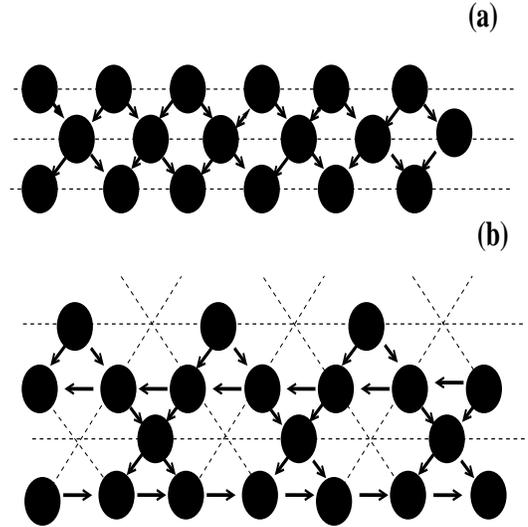}
\caption{High density liquid, \protect\( HDL\protect \) with density 1
(top) and low density liquid, \protect\( LDL\protect \) with density 3/4
(bottom)on the triangular lattice. The solid lines indicate the hydrogen
bonds where the arrows differentiate bond donors from bond acceptors.} 
\label{fig2} 
\end{center}
\end{figure}
\begin{figure}
\begin{center}
\includegraphics[height=4cm,width=6cm]{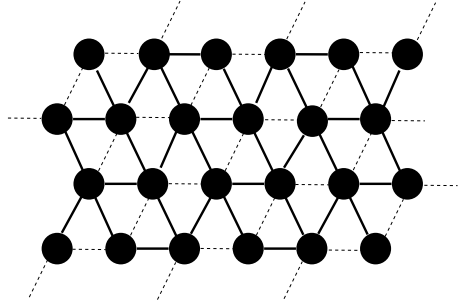}
\caption{A possible configuration if the system could have all sites 
distorted. The solid lines indicate the hydrogen bonds and the dashed 
lines are the non bonding interactions.}
\end{center}
\label{fig3}
\end{figure}

The model properties for finite temperatures were obtained through Monte Carlo
simulations in the grand-canonical ensemble using the Metropolis algorithm.
Some test runs were done for L=10, 20 and 50. A finite size
scaling analysis for the two critical temperatures show 
a small shift not relevant for our analysis.
Therefore, the 
detailed study of the model properties and the full phase diagrams was 
undertaken for an L=10 lattice. Runs were of the order of \( 10^{6} \) 
Monte Carlo steps. 

 We have considered the cases $u_{1}/v=1$ for proper bonds 
and $u_{2}/v=0.6,0.8$ 
for distorted bonds. 
Fig 4 illustrates the reduced temperature, $\bar{T}=k_BT/v$,  vs density coexistence phase diagram 
for $u_{2}/v=0.6$. The low
density liquid phase occurs in a very small interval of densities, related to
a very steep rise in the chemical potential isotherms. The line
of maximum densities (TMD) is also shown in the figure. 
\begin{figure}
\begin{center}
\includegraphics[height=7cm,width=9cm]{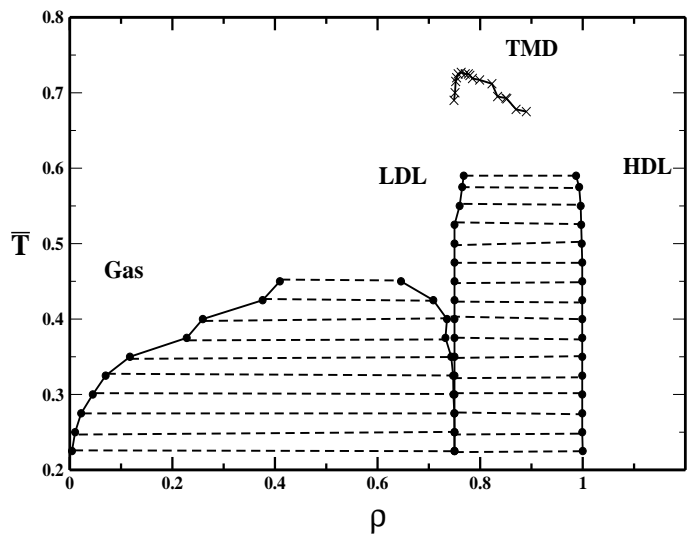}
\caption{Temperature vs. density coexistence phase-diagram for $u_{2}/v=0.6$.
The two liquid phases, the gas phase and the temperature of maximum density are illustrated.}
\end{center}
\label{fig4}
\end{figure}

Pressure 
was calculated by numerical integration of Gibbs Duhem equation 
at fixed temperature, from zero pressure at zero density. 
Figure 5 shows the $\overline{p}$-$\overline{T}$ ($\bar{p}=k_B p/v$) phase diagram for 
both $u_{2}/v=0.6$ and $u_{2}/v=0.8$. 
Data for the non-distorted version of the associating lattice gas
model \cite{He05a,He05b} is also shown for comparison.
The two coexistence lines, HDL-LDL and LDL-gas, and the TMD line, are 
present, but displaced,
if compared with the non-distorted case.

\begin{figure}
\begin{center}
\includegraphics[height=7cm,width=9cm]{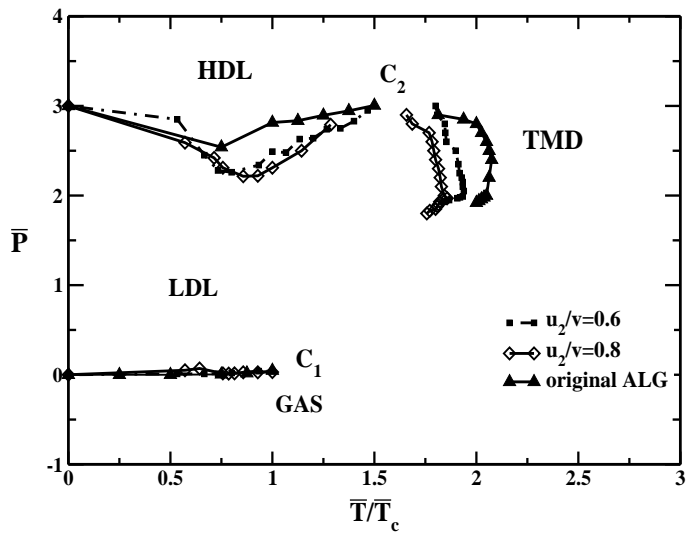}
\caption{Reduced pressure vs. reduced temperature for 
$u_{2}/v=0.6,0.8$. The temperatures are measured in 
terms of the LDL-gas critical point of each system. The original 
ALG
data  is also shown for comparison. The error bars are not shown
for clarity. }
\end{center}
\label{fig5}
\end{figure}
%
%
%
\begin{table}[hb]
\begin{center}
\begin{tabular}{|c|c|c|c|c|c|c|}
$System$&$\bar{T}_{c1}$&$\bar{\mu}_{c1}$&$\bar{P}_{c1}$&$\bar{T}_{c2}$& 
$\bar{\mu}_{c2}$&$\bar{P}_{c2}$\\ \hline \hline
$original ALG$& 0.45  & -2.05 & 0.1792 &0.65  &  1.71 & 3.005\\ \hline \hline
$u2/v=0.6$   & 0.375 & -2.09 & 0.04   &0.575 &  1.76 & 2.95 \\  \hline \hline
$u2/v=0.8$  & 0.375 & -2.09 & 0.03   &0.45  &  1.48 & 2.70 \\  \hline \hline
$symmetric$  & 0.55  & -1.86 & 0.07   &0.825 &  2.02 & 3.06 
\end{tabular}
\caption{Reduced temperature, chemical potential and pressure at the two critical points for the original model, distorted and symmetric models.}
\label{tab}
\end{center}
\end{table}
%
%

What is the effect of the hydrogen bond distortion? 
Both critical points for the model without distortions are at
higher temperatures when compared with the case in which distortions are 
allowed, as shown in Table I. 
The high degeneracy of
the distorted bond model smoothens the LDL-HDL transition thus destroying
the transition whereas it would still be present if the 
distortions were forbidden. 
The large degeneracy of the distorted bond molecules is also responsible 
for the larger slope
of the LDL-HDL coexistence line near the critical point.
This slope is positive, indicating, according to the Clayperon condition, a 
more
entropic LDL phase, if compared with the HDL phase. According to the same 
condition, the slope
of this line is proportional to the variation of entropy upon the change 
of phase. This implies that the
difference in entropy between the two phases is larger for the case of 
distorted bonds, implying a
higher entropy LDL phase for the distorted as compared with the 
non-distorted case.

From Table I we also observe that the critical temperatures for 
$u2/v=0.6$ are  closer than for $u2/v=0.8$  to the critical temperatures of the original
model. The case $u2/v=0.8$ represents a lower energy for the distorted 
bonds than the case  $u2/v=0.6$, consequently more distorted configurations
  are accepted for  $u2/v=0.8$ than for $u2/v=0.6$. 

\section{The symmetric-arms associating lattice gas  model}

In order to test the relevance of proton distribution entropy
with respect to the phase diagram properties, we study
a third version of our model. In this version we do not distinguish 
the acceptor and donor arms, as illustrated
in Fig 6. Distorted bonds are forbidden. Under this approach the model is
 considerably simplified and only 3 orientational states per particle remain.
\begin{figure}
\begin{center}
\includegraphics[height=6cm,width=8cm]{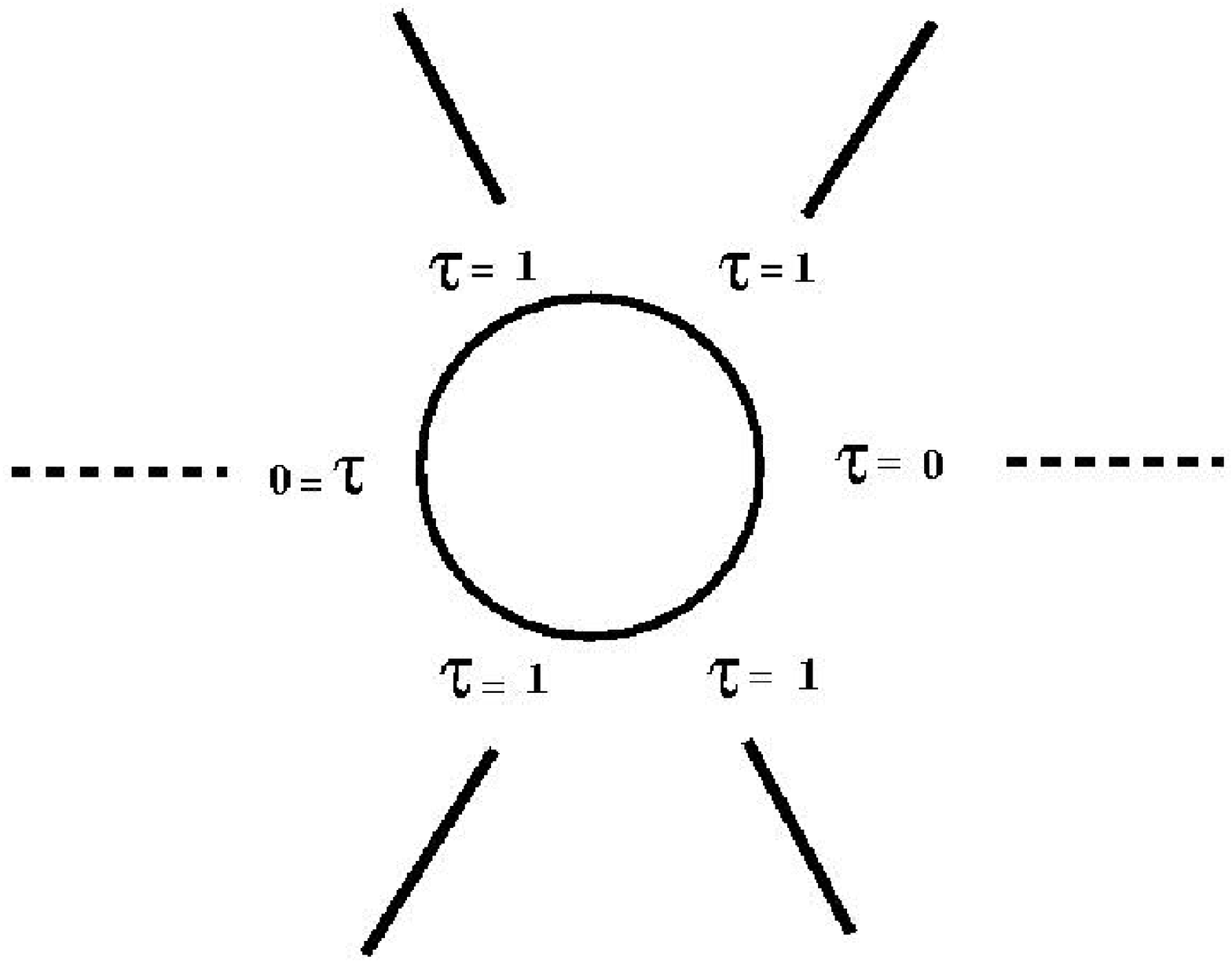}
\caption{The symmetric model}
\end{center}
\label{fi6}
\end{figure}
The overall energy is given by:
\begin{equation}
\label{E3}
E=(-v+2u)\sum _{(i,j)}\sigma _{i}\sigma _{j} 
+ u \sum _{(i,j)}\sigma _{i}\sigma _{j}\tau_{i}\tau_{j}
\end{equation}

The $\overline{p}$-$\overline{T}$ phase diagram for the symmetric-arms 
model has the same
structure of the phase diagram of the original associating lattice gas model. Figure 7 
shows the full 
$\overline{p}$-$\overline{T}$ coexistence lines and the TMD line. The
 two critical points,
at the ends of the gas-LDL and of the 
LDL-HDL coexistence lines are shown in comparison with
the other models in Table I.  Entropy effects here are
the opposite of those found for the distorted-bond model. Hydrogen 
bonding states have a
lower degeneracy, in comparison with the symmetric
 associating lattice gas model\cite{He05a,He05b} . Accordingly, the
 critical temperature
is increased and the slope
of the coexistence curve near the critical point diminishes (see Figure 7). 

\begin{figure}
\begin{center}
\includegraphics[height=7cm,width=9cm]{fig7.eps}
\caption{$\overline{p}$-$\overline{T}$ coexistence lines and 
the TMD line for symmetric model. The original ALG
data is also shown representing the asymmetric case.}
\end{center}
\label{fig7}
\end{figure}



\begin{figure}[htb]
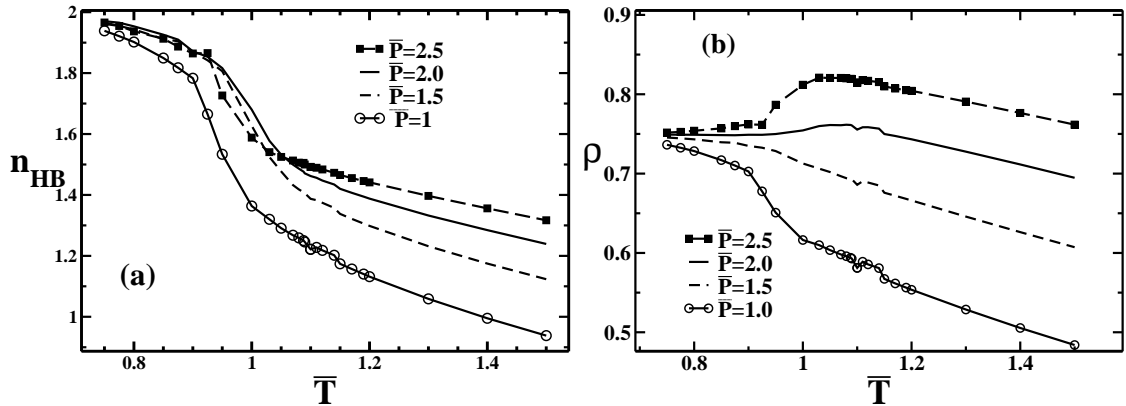

\begin{center}
\includegraphics[clip=true,scale=0.3]{fig8a.eps}
\vspace{0.5cm}
\includegraphics[clip=true,scale=0.3]{fig8b.eps}

\caption{(a) number of H-bond per particle vs. reduced temperature at fixed reduced pressures. (b) density vs. reduced temperature for fixed pressures.}
\end{center}
\label{fig16}
\end{figure}

\section{H-bond net disruption and density anomaly}
The lack of a proton entropy on bonds, in the case of the symmetric arms model, 
poses the question of 
its effect on the behaviour of those bonds,
under temperature and pressure variations. We have 
thus measured the number of hydrogen-bonds per 
particle over the whole phase diagram, and particularly in 
the region of the density anomaly. Figs 8a and
8b illustrate our findings. The two figures show: a) H-bonds per 
particle as a function of temperature;
and b) density as a function of temperature, at the same 
pressures. H-bond density decreases as temperature 
increases at all pressures. However, for those pressures for which a 
density anomaly is present (p=2 and 2.5),
a special behaviour of bonds can be seen: note the crossings of the 
isobaric H-bond densities. On the right-hand
side of the graph, h-bonds numbers increase steadily with pressure. On the 
left, at lower temperatures, h-bonds may decrease with pressure, as
indicated by the crossings. The crossing is a result of the fact that bond breaking
 is at a much higher rate for p=2.5 than for p=1.5.
This result implies a correlation between the 
behaviour of bonds and of density, in the anomalous region. 
At the pressures of anomalous density behaviour, a sharp decrease 
in bond density
 is seen as the density rises, at the lower temperatures. At the 
higher temperatures,
 and for normal density behaviour (which goes down with
temperature), h-bond densities decrease steadily with temperature,
at all pressures. An interesting picture
emerges, confirming and restricting an old qualitative prediction \cite{eisenberg}.
 The increase in density is associated
with the disruption of the hydrogen bond net. However, the \emph{rate} 
of disruption is important in order
to establish the presence of the density increment. If it is too 
small, the density anomaly is absent.

The coupling of anomalous density and anomalous bond disruption is
 not restricted to the symmetric arms model. The behaviour of hydrogen
 bonds for the other versions of the model, with either distorted bonds
 or assymetric
(donor-acceptor) arms, are entirely analogous (not shown).
In particular, this result implies that inspite of their presence, distortions
 may not play an essential role in relation to density anomaly.

\section{Conclusions}

In this paper we firstly address two questions: (i)what are the effects of
 distortions of the HB on the density anomaly or on  liquid polymorphism?
 (ii) is the hydrogen distribution entropy on the HB net relevant from 
the point of view of the characteristic features exhibited by these two properties? 

We have investigated these two questions in relation to an associating lattice-gas
model studied previously.  The original
model exhibits two liquid phases and a line of density anomalies (TMD). Two feature of
the phase diagram  merit some attention. 
One of them is the slope of the liquid-liquid coexistence line
which is positive except at low temperatures. This implies that the low density
phase has higher entropy than the high density phase, contrary to some previous expectations for
this line \cite{Po92}. 
A second feature is that the liquid-liquid 
critical temperature is higher than the gas-liquid critical temperature. Although present in
other models with liquid polymorphism (albeit without a density anomaly) \cite{Fra01}, this 
feature is contrary to what one would like for a model which describes water.

Could these two features be in any way modified by adding degrees of freedom, present in the
real system? Our answer to this question is no. Distortions do not reduce the liquid-liquid
critical temperature below the liquid-gas critical temperature and introduce more entropy
into the low density phase. 

On the other hand, removing
the distinction between donor and acceptor arms, responsible for proton
distribution entropy, implies reduction of the number of degrees of freedom
 and leads to opposite effects on the phase diagram. Our results show that these are the sole important
products of either modification: both the gas-liquid and the liquid-liquid critical temperatures and
also the slope of the liquid-liquid line,
are re-scaled, while the overall features of the phase-diagram remain unaltered.

A second point we were able to establish is the correlation between the rate of H-bond disruption and
the presence of a density anomaly. The latter is present only if the rate of bond breaking with increasing
temperature is sufficiently high. This property is independent of the presence of bond distortions or of
the distinction of acceptor and donor arms. This could be an indication of the greater relevance
 of bond disruption in relation to bond distortion in the arisal of a density anomaly.

\vspace*{1.00cm}

\noindent \textbf{\large Acknowledgments}{\large \par}

\vspace*{0.5cm} This work was supported by the Brazilian science agencies Capes, CNPq, FINEP and  Fapesp.

\end{document}